\documentclass[a4paper,12pt]{article}
\usepackage{amssymb,cite,graphicx}

\title{\bf Neutrino oscillation with flavor non-eigenstates and
CP-violating Majorana phases}

\author{\bf Rathin Adhikari \\
\normalsize Centre for Theoretical Physics, Jamia Millia Islamia Central
University\\ \normalsize Jamia Nagar, New Delhi 110025, India
\and \bf Palash B Pal\\
\normalsize Saha Institute of Nuclear Physics\\
\normalsize 1/AF Bidhan-Nagar, Kolkata 700064, India}
\date{December 2009}

\def\Eqn#1{Eq.\ (\ref{#1})}

\def\3Eqs#1#2#3{Eq.\ (\ref{#1}), (\ref{#2}) and (\ref{#3})}
\def\fig#1{Fig.~\ref{#1}}

\def\ket#1.{\left|#1\right>}
\def\mat#1{\mathbb #1}

\begin{document}

\maketitle

\begin{abstract}
  We analyze neutrino oscillation for the general case when the
  initial neutrino is not in a pure flavor state.  We show that, after
  such a neutrino beam propagates for a while, the probability of
  detecting any pure flavor state depends even on the CP-violating
  Majorana phases in the mixing matrix.  The dependence remains even
  when energy spectrum of the initial beam is taken into account.
  We discuss various implications of this dependence.
\end{abstract}

Through solar and atmospheric neutrino data as well as ground-based
oscillation experiments, we now know that neutrinos have mass, and
they mix \cite{review}.  Two of the three mixing angles are already
known to be non-zero, and in fact quite large, whereas for the other
there exists only an upper bound.  The masses themselves are not
known, and cannot be known from oscillation data alone, although we
have a good idea of the mass differences (more precisely, the
differences of mass-squared values) between different eigenstates.

Mass matrices contain information not only about mass eigenvalues and
mixing angles, but also about phases responsible for CP violation.
The number of such phases depend on whether the neutrinos are Dirac
fermions or Majorana fermions.  For $N$ generations of leptons, there
are $\frac12(N-1)(N-2)$ Dirac phases.  In case of Majorana neutrinos,
the number of phases is $\frac12N(N-1)$, i.e., $N-1$ more phases
compared to the Dirac case \cite{Bilenky:1980cx, Schechter:1980gr,
  Doi:1980yb}.  These extra CP-violating phases are sometimes called
``Majorana phases'', and we will use this terminology for the sake of
convenience.

An intriguing question is how to observe these phases should they
exist, i.e., if the neutrinos are Majorana particles.  Since Majorana
neutrinos are obtained in theories where lepton number is violated,
the observability of the Majorana phases was discussed in the context
of lepton number violating processes \cite{Schechter:1980gk,
  Doi:1980yb}, and for a long time it was believed that these phases
can be observed only in such processes.  Later, it was shown
\cite{Nieves:2001fc, Nieves:2002vq} that lepton number violation in
the process is not necessary in obtaining information about the
Majorana phases.

Inspired by this knowledge, we can ask the question whether the
Majorana phases can be observed in neutrino oscillation experiments.
This is what we try to do in this article, for neutrinos oscillating
in the vacuum and also in matter.

It is best to address this question assuming there are just two
generations of neutrinos, where the analysis can be carried out
analytically.  In this case, the mixing matrix contains only one phase
and it is of the Majorana type.  The flavor states $\nu_e$ and
$\nu_\mu$ are related to the mass eigenstates through the relation
\begin{eqnarray}
\pmatrix{\nu_e \cr \nu_\mu} = \pmatrix {\cos\theta & e^{i\alpha}
  \sin\theta \cr - e^{-i\alpha} \sin\theta & \cos\theta} 
\pmatrix{\nu_1 \cr \nu_2} \,.
\label{mixmat}
\end{eqnarray}
Suppose now we start with a monochromatic neutrino beam, with energy
$E$, which is not necessarily in one of the flavored states, but is a
general superposition of the two flavors:
\begin{eqnarray}
\ket \psi(0). &=& A \ket \nu_e . + B \ket \nu_\mu . 
\label{psi0}\\
&=& (A\cos\theta - Be^{-i\alpha}\sin\theta) \ket \nu_1. +
(Ae^{i\alpha}\sin\theta + B\cos\theta) \ket \nu_2. \,,
\end{eqnarray}
where $A$ and $B$ are real, and $A^2+B^2=1$.  We could have been more
general and assumed that there is a relative phase between $A$ and
$B$, but that is not necessary for the argument that we are going to
present.  Time evolution of this state gives, apart from an overall
phase that is unimportant, the result
\begin{eqnarray}
\ket \psi(t). = (A\cos\theta - Be^{-i\alpha}\sin\theta) \ket \nu_1. +
(A e^{i\alpha}\sin\theta + 
B\cos\theta) e^{-2i\delta} \ket \nu_2. \,,
\end{eqnarray}
where
\begin{eqnarray}
\delta = {m_2^2 - m_1^2 \over 4E} \; t \,,
\end{eqnarray}
$m_1$ and $m_2$ being the mass eigenvalues of $\nu_1$ and $\nu_2$.

We can now ask the question what is the probability of finding one of
the flavors, say $\nu_e$, in this resulting beam.  The answer is given
by 
\begin{eqnarray}
P_{\nu_e} (t) \equiv \left| \left< \nu_e | \psi(t) \right> \right|^2 \,.
\end{eqnarray}
Using \Eqn{mixmat} to express $\nu_e$ in terms of the eigenstates and
performing a few steps of trivial algebra, we obtain
\begin{eqnarray}
P_{\nu_e} (t) &=& A^2 (1 - \sin^2 2\theta \sin^2 \delta) +
B^2 \sin^2 2\theta \sin^2 \delta \nonumber\\*
&& - 2AB \sin 2\theta \sin \delta \left( \cos^2\theta \sin(\alpha +
        \delta) + \sin^2\theta \sin(\alpha - \delta)  \right) \,.
\label{monoenergy}
\end{eqnarray}
Clearly, if $A=1$, this will denote the survival probability of
$\nu_e$'s in a beam that started as $\nu_e$, and in this case our
expression reduces to the familiar expression for that case.  Our
result also agrees with the analysis of neutrino oscillation
probabilities performed \cite{Frieman:1987as} for $A$ and $B$ both
non-zero but assuming Dirac neutrinos, which implied $\alpha=0$.  The
same analysis can hold for Majorana neutrinos if CP violation is
neglected.  But our expression is more general than all these results,
and it shows that if neither $A$ nor $B$ vanishes, i.e., if the
initial beam is not in a pure flavor state, then the oscillation
probability depends on the Majorana phase $\alpha$.

What is more important, the dependence on $\alpha$ does not get washed
out even if we have a neutrino beam with a large spread of energy
which travels through a long enough duration of time.  In this case,
we need to take averages of the $\delta$-dependent quantities over the
energy spectrum.  Quantities like $\sin\delta$ or $\cos\delta$ vanish
under such averaging, but $\sin^2\delta$ averages to $\frac12$, and we
obtain 
\begin{eqnarray}
P_{\nu_e} (t) = A^2 (1 - \frac12 \sin^2 2\theta) +
\frac12 B^2 \sin^2 2\theta  - AB \sin 2\theta \cos 2\theta \cos\alpha
\,. 
\end{eqnarray}

We now proceed to derive the corresponding formulas for oscillations
in matter.  We denote the $2\times2$ Hamiltonian in the flavor basis by
\begin{eqnarray}
\mat H = \pmatrix{H_{11} & H_{12} \cr H_{21} & H_{22}} \,.
\end{eqnarray}
The Hamiltonian has to be hermitian, irrespective of whether we have
Dirac type of neutrinos or Majorana type, and so we must have $H_{11}$
and $H_{22}$  real and $H_{21}=H_{12}^*$.  Now, if there is a matrix
$\mat U$ of the form shown in \Eqn{mixmat} that makes $\mat U^\dagger
\mat H \mat U$ diagonal, the parameters of this matrix are given, in
terms of the elements of $\mat H$, by
\begin{eqnarray}
\theta = \frac12 \tan^{-1} \left( {2 \, |H_{12}| \over H_{22} - H_{11}}
\right) \,, 
\qquad
\alpha = \arg (H_{12}) \,.
\end{eqnarray}
For neutrinos passing through matter, the Hamiltonian depends on the
density \cite{Wolfenstein:1977ue}.  Density corrections appear in the
diagonal elements of the Hamiltonian written in the flavor basis.  As
a result, $\alpha$ does not change with density, although $\theta$
does.

So now, suppose the initial state produced is the superposition of
$\nu_e$ and $\nu_\mu$, as given in \Eqn{psi0}.  At the production
point, the density is such that the effective mixing angle is
$\theta_0$.  Thus,
\begin{eqnarray}
\pmatrix{\nu_e \cr \nu_\mu} = \pmatrix {\cos\theta_0 & 
  e^{i\alpha} \sin\theta_0 \cr - e^{-i\alpha} \sin\theta_0 & \cos\theta_0} 
\pmatrix{\nu_1 \cr \nu_2}_0 \,,
\end{eqnarray}
where the subscript ``0'' indicates the production point in matter.
Thus, 
\begin{eqnarray}
\ket \psi(0). = (A\cos\theta_0 - Be^{-i\alpha}\sin\theta_0) \ket
\nu_1._0 + (Ae^{i\alpha}\sin\theta_0 + B\cos\theta_0) \ket \nu_2._0 \,. 
\end{eqnarray}

The probability that this state is in the first eigenstate is given by
$|A\cos\theta_0 - B\sin\theta_0e^{-i\alpha}|^2$.  In the adiabatic
case, this probability remains constant, although the eigenstate, as a
superposition of the flavor states, changes with density.  If at the
end of the journey the neutrino is detected in the vacuum where
\Eqn{mixmat} holds, the probability of finding a $\ket\nu_e.$ in this
state is just $\cos^2\theta$.  Similarly adding the contribution from
the other eigenstate, we obtain
\begin{eqnarray}
P_{\nu_e}^{\rm (ad)} (t) &=& |A\cos\theta_0 - B\sin\theta_0e^{-i\alpha}|^2
\cos^2\theta + 
|A\sin\theta_0 e^{i\alpha}+ B\cos\theta_0|^2 \sin^2\theta \nonumber\\*
&=& \frac12 A^2 \left( 1 + \cos 2\theta_0 \cos 2\theta \right)
+ \frac12 B^2 \left( 1 - \cos 2\theta_0 \cos 2\theta \right)
\nonumber\\* 
&& - AB \sin 2\theta_0 \cos 2\theta \cos\alpha \,.
\end{eqnarray}
The superscript `ad' reminds us that this is the result if the
adiabatic condition, described above, holds.  This is, of course, the
energy-averaged expression, because we have neglected the oscillatory
terms by working with an explanation in terms of probabilities
\cite{Barger:1986ww} and not of amplitudes.  But even in this
expression, there is an $\alpha$-dependence.

If the conditions are non-adiabatic, then neutrinos can jump from one
eigenstate to another.  If this jump probability is called $P_{\rm
  jump}$, then the probability of finding $\nu_e$ would be given
by~\cite{Haxton:1986dm,Parke:1986jy}
\begin{eqnarray}
P_{\nu_e} = (1-P_{\rm jump}) P_{\nu_e}^{\rm (ad)} + P_{\rm jump}
(1-P_{\nu_e}^{\rm (ad)}) \,. 
\end{eqnarray}
The jumping probability depends on the expression of the eigenvalues
in terms of the parameters that appear in the Hamiltonian in the
flavor basis.  These eigenvalues are independent of $\alpha$ and
therefore the jumping probability is independent of $\alpha$ as well.
However, the expression for $P_{\nu_e}$ still depends on $\alpha$
through the adiabatic probability that appears in this expression.

There are claims in the literature that the Majorana phases are
unobservable in neutrino oscillation experiments, whether in vacuum
\cite{Bilenky:1980cx, Schechter:1980gr, Doi:1980yb} or in matter
\cite{Langacker:1986jv}.  However, these claims have all been made in
the context of initial neutrino states of pure flavor.  This is
consistent with our results since the $\alpha$-dependence in all
relevant formulas appear with a co-efficient $AB$, and therefore
vanish for pure flavor states which have either $A=0$ or $B=0$.

Presence of a non-zero value of the CP-violating phase $\alpha$ can
introduce some qualitative features in neutrino oscillation
probabilities.  As an example, suppose we have an initial neutrino
beam with some known admixture of two flavors.  If $\alpha=0$, the
probability of finding $\nu_e$ after this beam travels for a time
$t$ in the vacuum is given by
\begin{eqnarray}
P_{\nu_e} (t) \bigg|_{\alpha=0} = A^2 - \Big[ (A^2-B^2) \sin^2 2\theta
+ 2AB \sin 2\theta \cos2\theta \Big] \sin^2 \delta \,. 
\end{eqnarray}
If the initial state was such that both $A$ and $B$ are positive with
$A>B$, the right hand side of this expression is always less than
$A^2$, assuming $\theta<\pi/4$ which is certainly valid for
$\nu_e$-$\nu_\mu$ oscillation, as inferred from various solar and
terrestrial experiments \cite{review}.  However, if $\alpha\not=0$, we
need to use the expression given in \Eqn{monoenergy}, and $P_{\nu_e}
(t)$ is no longer guaranteed to be less than $A^2$.  Thus, if one
observes a reinforcement of the dominant component of the beam in an
oscillation experiment, it definitely signals a non-zero value of the
phase $\alpha$.

\begin{figure}

\centerline{\input{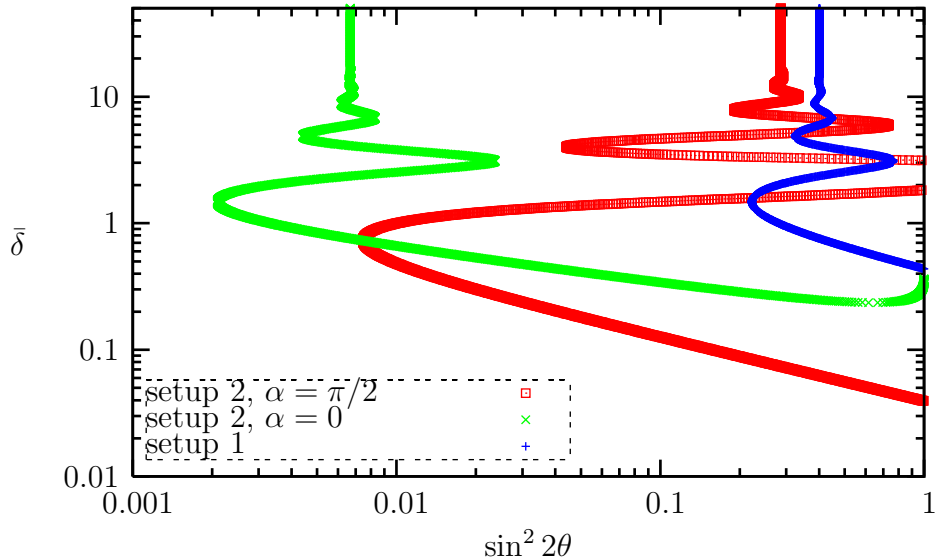}}

  \caption{Various equal probability contours.
    Details have been given in the
    text.}\label{f:eqprob}
  
\end{figure}

Let us conclude this article with a qualitative feeling for the
magnitude of the importance of $\alpha$ oscillation formulas, which
might help in devising experimental techniques for finding $\alpha$
from neutrino oscillation experiments.  We will do this by considering
fictitious data coming from two different oscillation experiments.  It
has been argued \cite{Pal:1998qv} that the analysis of data from a
single oscillation experiment is conveniently done by introducing
the dimensionless parameter
\begin{eqnarray}
\bar\delta \equiv {m_2^2 - m_1^2 \over 4 \langle E \rangle} \; t \,,
\end{eqnarray}
where $\langle E \rangle$ is the average energy in the incoming beam.
Different experiments can have different average energy in the
incoming beam and different distances between the source and the
detector, and consequently different values of $\bar\delta$ even when
they are measuring oscillation between the same two flavors.  In order
to provide an illustrative example of the point we are discussing
here, we consider, for the sake of simplicity, that two experiments
have the same $\langle E \rangle$ and the same source-detector
distance (or at least the same ratio of the two quantities just
mentioned), and therefore the same value of $\bar\delta$.  One of
these experiments is working with an initial beam that is purely
$\nu_e$, and observes that $P_{\nu_e}=0.8$ at the detection point.  We
can ask which values of $\bar\delta$ and the mixing angle can make it
possible.  The answer will be independent of $\alpha$ since the
initial beam was purely $\nu_e$, and has been shown in \fig{f:eqprob}.
The shape of this line of course depends on the energy spectrum of the
initial beam.  For the sake of definiteness, we have taken a Gaussian
energy spectrum with a standard deviation equal to $0.2\langle E
\rangle$.

Now suppose that in the second experiment, the initial neutrino beam
has $A=0.8$ and $B=0.6$ in the initial beam, and at the detector, one
finds $P_{\nu_e}=0.6$.  The energy spectrum is the same.  If
$\alpha=0$, the equal probability contour is given by the line that
reaches leftmost in \fig{f:eqprob}.  It is worth pointing out a
special feature of this line compared to the line obtained from the
first experiment.  Near the bottom end of the line, we have two
solutions of the mixing angle for the same value of $\bar\delta$.
This happens because, when $A$ and $B$ are both non-zero, the
$\theta$-dependence of the expression in \Eqn{monoenergy} is not
monotonic.

Anyway, this line is clearly inconsistent with the result of the first
experiment.  However, if $\alpha\not=0$, this need not be the case.
We show the contour for $\alpha=\pi/2$ in \fig{f:eqprob} with a thick
solid line, which has clearly some intersections with the curve from
the first experiment.  Thus, non-zero values of $\alpha$ will make the
results of the two experiments consistent with each other, and the
value of $\alpha$ can even be determined from such data from two
experiments. 

Real experiments, of course, will have some error bar on the detected
probability, so there will be a band rather than a contour
corresponding to a single experiment.  There will be other
complications in the analysis because the energy spectra of no two
experiments will be the same.  Obtaining an initial beam which is not
a flavor eigenstate is not straight forward as well.  Here, we have
only discussed the matters of principle and indicated possible ways in
which the CP-violating phase $\alpha$ might be detected from neutrino
oscillation data.

\noindent ---------------------

PBP wants to thank the hospitality of the Instituto Superior Tecnico,
Lisboa, where part of the work was done.

\end{document}